# FOCUSING MIRROR WITH TUNABLE ECCENTRICITY

Moritz Stürmer, Matthias C. Wapler, Jens Brunne and Ulrike Wallrabe
University of Freiburg – IMTEK, Germany
E-mail: wallrabe@imtek.uni-freiburg.de

## ABSTRACT

We present a new kind of varifocal mirror with independently adjustable curvatures in the major directions. For actuation we use two stacked piezo bending actuators with crossed in-plane polarization. This mirror can be used for example as an off-axis focusing device with tunable focal length and compensation for a variable angle of incidence or for coma correction. We demonstrate the prototype of such a mirror and characterize the mechanical deflection, as well as the focusing capabilities.

## INTRODUCTION

Common tunable focusing mirrors, e.g. [1], can only be used parallel to the optical axis since their surfaces form a rotationally symmetric paraboloid, and off-axis illumination leads to distortion. We show a deformable mirror which overcomes this restriction based on a longitudinal mode piezo bending actuator, which combines two hyperbolic deflections. Hence, its resulting deflection is an elliptic paraboloid which features different adjustable curvatures along its major directions. In contrast to other known deformable mirrors that use the transverse piezo effect, e.g. [2], our device can be operated with just two different voltages.

## WORKING PRINCIPLE

Piezo bending actuators can be operated in the transverse or the longitudinal mode. While the transverse mode is usually favored due to simple planar electrodes, the longitudinal actuators' key feature is usually considered to be the higher electro-mechanical coupling constant $d_{33} \approx -2\, d_{31}$.

In-plane polarized piezo actuators as they have been investigated, e.g., in [3], can be realized using interdigitated electrode (IDE) geometries as depicted in Fig. 1a. An elongation perpendicular to the electrode fingers is caused, if these actuators are driven in the polarization direction. Since there is also the transverse piezo effect with a negative coupling constant, there is always a negative strain in direction of the finger electrodes. Therefore, such an actuator yields a hyperbolic paraboloid as surface profile, i.e., it forms a saddle. The latter is true for a piezo actuator which is firmly connected to a passive layer.

Our new concept consists of three layers (Fig. 1b): two piezo bending actuators, whose finger structures are perpendicularly oriented, and one mechanically passive mirrored glass layer. Since the stiffness of this passive layer is much higher than the one of the piezos, the neutral plane of the deflection is inside the glass layer. This means that the direction of the out-of-plane deflection is governed by the passive layer.

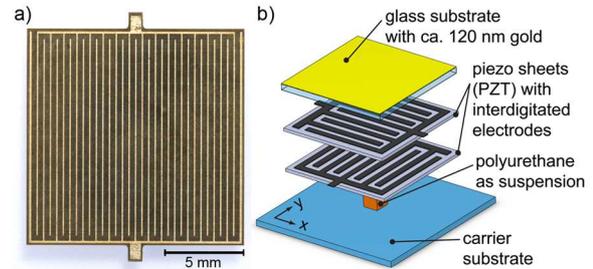

Fig. 1: a) Photograph of the IDE structure on one of the actuators; b) Schematic layer stack (not to scale).

As the electrodes are spread over the whole piezo actuator and edge effects are small the full area of approx. 14 x 14 mm is deflected and can be used as a mirror.

## FABRICATION

To demonstrate the working principle we fabricated a mirror device with integrated actuator. The piezo actuators are commercially available 120 µm thick PZT films with silver electrodes which are depolarized by heat (20 min @ 420°C) prior to further processing. The electrodes on both sides are structured with a UV-laser. A spacing of 320 µm between the single electrode fingers is favored due to a trade-off in different aspects of field homogeneity. The lower piezo was thinned down to 75 µm to move the neutral plane of the deformation into the glass layer. A successive $HNO_3$ cleaning dip enhanced by ultrasound ensures that all remaining silver particles are removed from the surface. For polarisation an electric field of 2 kV/mm is applied to the structured electrodes for 10 min to allow for sufficient time for creep. The electrodes are spin coated with a 10 µm thin layer of SU-8 photoresist. Together with the epoxy glue which is used to firmly bond the two piezo layers together it prevents contact between the electrodes. Finally, the piezo actuators are bonded to a 200 µm thick gold-plated glass substrate using epoxy glue. Boundary effects are avoided by freely suspending the mirror device on a piece of soft elastic polyurethane material (8 $mm^3$) at the centre.

## CHARACTERIZATION

The mechanical deflection of the device was measured using a laser distance sensor with which the surface of the mirror was scanned while the actuators were driven with the supply voltages. 120 V was chosen as a conservative maximum voltage to avoid damage.

To evaluate the bending radii in the x- and y-direction of the mirror a parabolic function was fitted to the surface deflection for both directions. The coefficient of the quadratic term corresponds to the inverse bending radius of the surface. In a first experiment the dependence of curvature on the applied voltages was determined by actuating each piezo layer separately. Assuming that the system behaves linearly, it was found that the voltage ratio of upper to lower piezo for a unidirectional strain in x- and y- direction must be 1.86 and 0.58, respectively. For a spherical deflection it must be 0.98. These values were then applied to the slopes of the driving voltages shown in the upper part of Fig. 2 to demonstrate the tunability of the surface. The driving frequency was kept low to avoid any resonance.

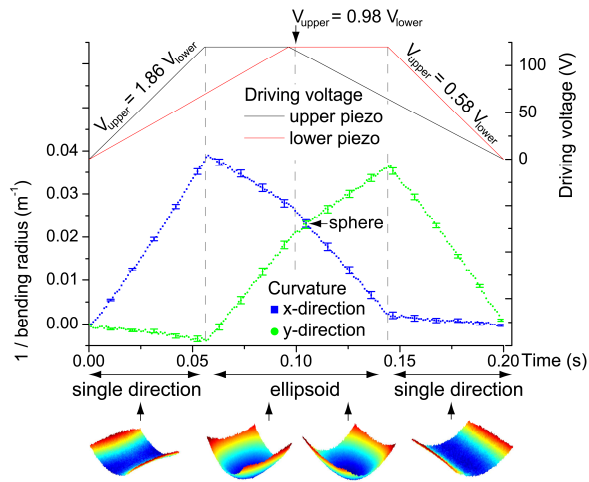

Fig. 2: Top: Actuation voltages, middle: the curvatures in x- and y-direction, the exemplarily given error bars indicate the fit-parameter standard error; bottom: visualization of the scanned surface profile.

One can see that the x- and y-bending radii can be controlled independently leading to ellipsoidal or even to almost unidirectionally bent surfaces. The linear assumption works well, and crosstalk between the layers is small. In terms of focal length of the mirrored passive layer this means that values of approx. 12.5 m to ∞ can be realized in each direction.

The actual focusing performance is determined with a simple optical setup. The mirror device is illuminated with a slightly defocused laser beam of ca. 4 mm diameter under an incidence angle of 45°. The spot is monitored with a camera in a fixed distance to the mirror of ca. 2.9 m while the driving voltages are varied from 60 to 120 V for each piezo in steps of 3 V. The full width at half maximum of the spot dimensions in the major camera directions is then determined by a Gaussian fit of the intensity data. The spot diameter is calculated assuming that the spot has an elliptical shape. Fig. 3 depicts the spot diameter in dependence of the applied voltages. The circles indicate the position of the smallest spot sizes observed. It becomes apparent that the minimum spot size is shifted towards voltage ratios which lead to a by $\sqrt{2}$ stronger deflection in the tilted direction of the mirror. This means that ellipsoidal deflection decreases spot sizes and hence improves the focusing compared to a simple spherical deflection.

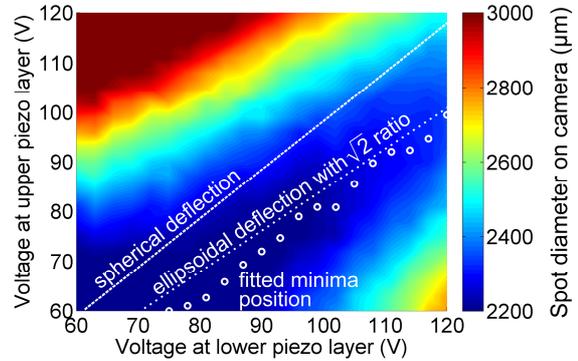

Fig. 3: Focusing behavior depending on voltage ratio: lines represent theoretical ratios for spherical and ellipsoidal deflections, circles minimum spot sizes.

## CONCLUSIONS

We have verified a principle of a varifocal mirror with tunable eccentricity based on a 3-layer piezo composite. The mechanical characterization of a prototype showed a tuning range from a spherical surface up to differently oriented unidirectional profiles. Its measured mechanical and optical properties confirm that this mirror can be used for off-axis focusing. The focal range is still in the region of 10 m, but can be improved, for example, by using thinner substrates and different materials.

## ACKNOWLEDGEMENTS

This research was financed by the German Research Foundation (DFG) and by the Baden-Württemberg Stiftung gGmbH.

## REFERENCES

[1] M. Mescher, M. L. Vladimer, J. J. Bernstein, "A novel high-speed piezoelectric deformable varifocal mirror for optical applications" Proc. 15th IEEE Int. Conf. Micro Electro Mechanical Systems, 2002, pp. 511–515.
[2] I. Kanno, T. Kunisawa, T. Suzuki, H. Kotera, "Development of Deformable Mirror Composed of Piezoelectric Thin Films for Adaptive Optics" IEEE Journal of Selected Topics in Quantum Electronics, vol. 13, no. 2, pp. 155–161, 2007.
[3] W. Beckert, W. S. Kreher, "Modelling piezoelectric modules with interdigitated electrode structures" Computational Materials Science, vol. 26, pp. 36–45, Jan. 2003.